\documentclass[12pt, a4paper]{article}
\usepackage[utf8]{inputenc}
\usepackage[english]{babel} 
\usepackage{mathptmx} 
\usepackage{amsmath}
\usepackage{amssymb} 
\usepackage{graphicx}
\usepackage[left=2.5cm, right=2.5cm, top=2.5cm, bottom=2.5cm]{geometry}
\usepackage{setspace}
\usepackage{booktabs}  
\usepackage{array}        
\usepackage{adjustbox}    
\usepackage{longtable}    
\usepackage{tabularx}      
\usepackage{ragged2e}

\usepackage{hyperref}
\hypersetup{
  colorlinks=true,
  linkcolor=blue,
  urlcolor=cyan,
  citecolor=blue
}
\usepackage{natbib}
\newcolumntype{L}[1]{>{\RaggedRight\arraybackslash}p{#1}} 
\newcolumntype{Y}{>{\centering\arraybackslash}X}

\title{\textbf{From Murals to Memes: A Theory of Aesthetic Asymmetry in Political Mobilization}}
\author{Ricardo Alonzo Fernández Salguero}
\date{\today}

\begin{document}

\maketitle

\begin{abstract}
\noindent Why have left-wing movements historically integrated participatory art forms—such as murals and protest songs—into their praxis, while right-wing movements have prioritized strategic communication and, more recently, the digital culture of memes? This article introduces and develops the concept of \textbf{aesthetic asymmetry} to explain this fundamental divergence in political action. We argue that this asymmetry is not coincidental but rather the result of four interconnected structural factors: the organizational ecosystem, the moral-emotional framework, the material supports, and the historical tradition of each political spectrum. While the left employs art in a \textit{constitutive} manner to forge community, solidarity, and hope, the contemporary right uses it \textit{instrumentally} to mobilize polarizing affects like humor and resentment. Analyzing comparative literature, from the Theatre of the Oppressed to the meme wars of the alt-right, we nuance this distinction and demonstrate how the aesthetic logic of each pole corresponds to its strategic objectives. The article culminates by proposing a \textbf{prescriptive model for artistic action}, synthesizing the keys to effective mobilization into emotional, narrative, and formatting strategies. We contend that understanding this asymmetry is crucial for analyzing political communication and designing cultural interventions capable of generating profound social change.
\end{abstract}

\section{Introduction}

At one end of the political spectrum, the guitar of Víctor Jara resonates, a symbol of popular song as a tool for consciousness and unity. At the other, an ironic and viral meme proliferates, designed to ridicule an adversary and consolidate a group identity within the digital ecosystem. This visual and sonic polarity encapsulates a fundamental puzzle for political science and the sociology of social movements: why is there such a marked asymmetry in the use of art between the left and the right? While the literature has abundantly explored the relationship between art and politics, the \textit{systematic differences} in how each ideological spectrum conceives of and deploys aesthetic tools remain notably under-theorized.

This article directly addresses this gap by introducing and developing the concept of \textbf{aesthetic asymmetry}. We maintain that the historical affinity of the left for organic and participatory art (muralism, community theater, protest songs) and the preference of the right for strategic communication and, today, digital aesthetics (memes, conservative pop culture) are not mere stylistic choices. They are, on the contrary, the result of deep structural divergences. Our argument is built on four interconnected factors: \textbf{1) the organizational ecosystem}, which on the left requires building community from the grassroots; \textbf{2) the moral-emotional framework}, which mobilizes hope against injustice; \textbf{3) the material supports}, which prioritize the gathering of bodies in public space; and \textbf{4) the accumulated aesthetic tradition}, which conceives of art as a pedagogical practice.

Throughout this analysis, we contrast these dynamics with the strategies of the right, examining both its traditional forms of communication and its successful foray into the digital culture war, a phenomenon that compels a nuancing of any simplistic distinction. Finally, the article is not limited to diagnosis; it distills these lessons into a practical model for designing effective artistic interventions. In doing so, we offer a unified framework for understanding how the synergy between art, emotion, and strategy becomes the engine of political mobilization in an increasingly polarized world.

\section{Structural Factors of Asymmetry: Ecosystem, Moral Framework, and Tradition}

The historical preference of left-wing movements for organic and participatory art, in contrast to the right's focus on strategic communication, is grounded in four interconnected pillars that define the nature of their political action.

The first factor is the \textbf{organizational ecosystem and purpose}. Historically, the left has built its power from the social base, through a dense network of trade unions, mass parties, popular athenaeums, cultural `peñas', and community festivals. In these spaces, art is not an accessory but an integral part of social and organizational life. Theatre for Development (TfD) in Africa, for example, emerges as a practice that seeks to empower communities from within their own cultural forms to address concrete problems, from public health to local governance \citep{Balme_Hakib_2023}. Similarly, the \textit{son jarocho} in Mexico is revitalized as a community practice that opposes colonial and capitalist discourses, creating a ``Black sense of place'' that is both aesthetic and political \citep{Astorga_de_Ita_2024}. These artistic ecosystems not only disseminate a message but also generate cohesion, resilience, and social capital. In contrast, the right has traditionally prioritized institutional apparatuses such as think tanks, lobby groups, and control of mass media, where the objective is not community-building from below but vertical persuasion and agenda-setting.

The second pillar is the \textbf{moral framework and emotional structure}. Left-wing movements often build their narratives around a dramatic arc that moves from denouncing \textbf{injustice} to building \textbf{solidarity} to achieve the \textbf{hope} of a better future. This narrative requires art to be felt and embodied. Music, as studies on South African `freedom songs' show, allows for the collective expression of pain and anger but also fosters resilience and democratic participation through practices like call-and-response \citep{Jolaosho_2019}. Art enables the transformation of individual suffering into a collective force. Conversely, conservative narratives often focus on defending \textbf{order} against a \textbf{threat} (external or internal), which requires an act of \textbf{defense}. This framework is better suited to brief, direct, and disciplined messages, typical of strategic communication and political marketing, which seek to activate emotions like fear or national pride in a more instrumental way.

The third factor is \textbf{material supports and corporeality}. The community-based art of the left—murals, street theater, collective singing—convenes and organizes \textbf{bodies in public space}. It is an inherently social and physical practice that builds community in the very act of its creation. Participatory artistic practices, as studied in contexts of civic activism, generate social capital, strengthen interpersonal bonds, and encourage participation in community life \citep{LeRoux_Bernadska_2014, NORC_2021}. The political marketing of the right, on the other hand, optimizes the dissemination of \textbf{messages through mass media} (TV spots, slogans, and today, viral memes). Its logic is not to bring bodies together but to maximize the reach and repetition of the message. The effectiveness of right-wing memes, for example, lies not in collective deliberation but in their ability to polarize and mobilize affects through humor and repetition on digital networks \citep{Schmid_2025}.

The \textbf{pedagogical and aesthetic tradition} also plays a key role. From Mexican muralists like Rivera and Siqueiros, through the Chilean New Song of Víctor Jara and Quilapayún \citep{McSherry_2017}, to Augusto Boal's Theatre of the Oppressed, there is a solid line of continuity on the left that conceives of art as a tool for consciousness-raising and popular pedagogy. The artist is both a creator and an educator, and the audience is an active participant in the construction of meaning. This tradition has generated a vast repertoire of methodologies and practices, such as participatory theater for development, which has been institutionalized in universities and NGOs worldwide \citep{Balme_Hakib_2023}. The right, while having its own aesthetic traditions (associated with order, nation, or religion), has not developed a body of participatory and pedagogical artistic practices with the same coherence and historical continuity.

\section{Exceptions and the Contemporary Aesthetic Reconfiguration on the Right}

Although the historical asymmetry is evident, it would be a mistake to ignore counterexamples and recent transformations, especially in the context of digital culture. The populist and alternative right has demonstrated a remarkable ability to appropriate aesthetic and emotional tools, subverting the idea that mobilizing art is an exclusive heritage of the left.

The most notable phenomenon is the \textbf{meme wars and conservative digital culture}. Memes, as humorous and viral visual artifacts, have become a powerful tool for political communication. Research on the use of memes in far-right Telegram channels shows that humor is used strategically to mask extremist content, increase its reach, and normalize radical ideologies \citep{Schmid_2025}. These memes function as a ``Trojan horse'': their playful appearance and their ability to generate group cohesion through the mockery of the `other' (the establishment, progressives, minorities) make them highly effective for mobilizing emotions and polarizing the debate. A longitudinal experiment on the impact of political memes on Facebook revealed that exposure to memes incongruent with one's own ideology (e.g., a left-wing voter seeing right-wing memes) can intensify polarization, especially among stronger partisans \citep{Galipeau_2023}. This suggests that the aesthetics of the meme do not so much persuade as they reinforce and entrench pre-existing identities, a key mechanism in the strategy of the contemporary right.

In addition to digital culture, there are more traditional forms of cultural production with a conservative bias. \textbf{Film and music with an emphasis on faith, fatherland, and family} have demonstrated their ability to reach mass audiences, often operating through well-established commercial markets. These artistic productions, though not always framed as `activism', serve a political function by reinforcing a set of conservative values and building a shared cultural identity. Likewise, the \textbf{aesthetics of order and national pride}—manifested in military parades, patriotic anthems, and the massive use of national symbols—is a form of political performance that mobilizes powerful emotions of belonging and loyalty. Although not classified as `protest art', its function is eminently political and affective.

These counterexamples do not invalidate the thesis of historical asymmetry, but they do nuance it. They show that the right is adapting its communication strategies to incorporate aesthetic and emotional tools, albeit with a different logic. While left-wing art often seeks to create spaces for deliberation and community participation, the aesthetics of the populist right focus on virality, polarization, and the reinforcement of group identity against a threatening `other'. Both use emotion, but for different ends and through different mechanisms.

\section{Analyzing Effectiveness: A Practical Model for Emotional Mobilization}

Based on the accumulated evidence on the impact of art and communication in social movements, it is possible to distill a practical model for designing interventions that seek to ``persuade by transferring emotions.'' This approach is not exclusive to one ideology, but its emphasis on empathy, solidarity, and community building aligns with the historical tradition of the left. The effectiveness of these practices can be analyzed through their emotional, narrative, symbolic, and formatting components. The following tables break down this model, integrating findings from various studies.

Political communication does not operate in a vacuum. Direct persuasion by leaders, for example, has been shown to be effective not only in changing attitudes on specific issues but also in modifying perceptions of the leader and the electoral behavior of citizens, as observed in field experiments with members of the U.S. Congress \citep{Minozzi_2015}. Art and performance can function as a form of ``distributed leadership,'' where the work itself performs this persuasive function. The ``emotional cocktail'' (Table \ref{tab:emociones}) is the basis of this process. Research on creative activism confirms that interventions generating positive emotions like humor or surprise capture more attention and are remembered more easily than conventional forms \citep{Duncombe_Harrebye_2022}. However, moral anger at a concrete injustice remains a fundamental driver of collective action.

\begin{table}[ht]
\centering
\caption{The Emotional Cocktail of Mobilization}
\label{tab:emociones}
\begin{tabularx}{\textwidth}{L{4cm} X}
\toprule
\textbf{Emotion} & \textbf{Function in Mobilization and Supporting Evidence} \\
\midrule
\textbf{Moral Anger} & Catalyzes action against an injustice perceived as concrete and intolerable. It is the primary engine of discontent. Theatrical campaigns against HIV/AIDS stigmatization in Uganda, for example, channel this anger into constructive community action \citep{Lee_2022}. \\
\textbf{Tenderness \& Care} & Fosters solidarity and mutual care, humanizing the struggle and creating affective bonds. It counteracts cynicism. Participatory art has been shown to improve resilience and well-being in communities with chronic medical conditions, strengthening the social fabric \citep{Gallagher_2024}. \\
\textbf{Pride \& Dignity} & Reinforces the identity and value of the collective (of a trade, a territory, a culture). It is fundamental for self-affirmation. The \textit{son jarocho} is an example of how music can restore the dignity of historically marginalized communities \citep{Astorga_de_Ita_2024}. \\
\textbf{Hope} & Articulates a vision of a possible victory, however small. It is the antidote to resignation. Protest songs, like those of the Nueva Canción, often end with a call to action based on the certainty of a better future \citep{McSherry_2017}. \\
\textbf{Humor \& Irony} & Disarms the spectator's defenses, criticizes power non-confrontationally, and generates a collective pleasure that fosters cohesion. The strategic use of humor in right-wing memes demonstrates its power to mask radical content and increase its reach \citep{Schmid_2025}. \\
\bottomrule
\end{tabularx}
\end{table}

Once activated, emotions must be channeled through a coherent narrative. A minimal story, as detailed in Table \ref{tab:narrativa}, structures the experience and orients it toward action. This narrative model is more effective when it relies on local symbols and motifs, as observed in community theater, which uses the ``idiom, language, and cultural forms of the people'' to address their problems \citep{Idebe_2023}. The rhythm of a \textit{cumbia villera} or the iconography of a neighborhood mural are, in themselves, political arguments.

\begin{table}[ht]
\centering
\caption{Narrative Model for Collective Action}
\label{tab:narrativa}
\begin{tabularx}{\textwidth}{l X}
\toprule
\textbf{Step} & \textbf{Description and Function} \\
\midrule
1. \textbf{The Hook} & A powerful and concrete image, sound, or scene that captures immediate attention. Example: ``a stalk of wheat as a flag'' in a sugarcane field. It functions as the story's sensory anchor. \\
2. \textbf{The Person} & Introduce an individual with a name, face, and occupation. Do not speak of ``the workers,'' but of ``Alba, a sugarcane cutter.'' This generates empathy and allows for identification. Research on political persuasion often focuses on how messages affect specific individuals \citep{Fisman_2025}. \\
3. \textbf{The Obstacle} & Define the problem in a visible and tangible way: an unfair price, an unpayable debt, a service that never arrives. The obstacle must be understandable and perceived as surmountable through collective action. \\
4. \textbf{The Community} & Show the individual acting with others: an assembly, a `fandango', a union, a group of neighbors. This transforms an individual complaint into a collective demand and shows that the solution is social. Participatory art has been shown to be a catalyst for civic engagement \citep{Ho_2021}. \\
5. \textbf{The Micro-Victory} & Narrate a concrete change, however small: a meeting secured, a petition delivered, a small fund raised. This generates a sense of efficacy and shows that action has real consequences, feeding hope. \\
6. \textbf{The Call to Action} & Conclude with a clear, simple, and achievable short-term call to action. What can the person who has been emotionally mobilized do tomorrow? The action must be the logical next step of the narrative. \\
\bottomrule
\end{tabularx}
\end{table}

Formats and dissemination strategies are equally important. As summarized in Table \ref{tab:formatos}, practices that convene bodies and emotions, such as participatory murals or collective singing, have a profound impact on group cohesion. Theatre for Development in Africa has repeatedly shown that active participation in artistic creation is key to message appropriation and mobilization \citep{Banda_Mpolomoka_2023}. Furthermore, in the digital age, it is essential to design art to be shared, not just consumed. Simple choreographies, catchy choruses, or meme templates are invitations for others to appropriate the work and make it their own. Finally, impact evaluation (Table \ref{tab:metricas}) must go beyond vanity metrics like `likes' and focus on observable behavioral changes and the sustainability of collective action, always maintaining an ethics of emotion that avoids manipulation and respects the dignity of all actors, including adversaries.

\begin{table}[ht!]
\centering
\caption{Components for an Aesthetics of Persuasion: Format and Dissemination}
\label{tab:formatos}
\begin{tabularx}{\textwidth}{L{4.5cm} X}
\toprule
\textbf{Component} & \textbf{Description and Examples} \\
\midrule
\textbf{Formats that Convene Body and Emotion} & Prioritize practices that bring people together in physical space and actively involve them.
\begin{itemize}
    \item \textbf{Participatory Mural}: A base sketch is designed, but it is completed in open workshops with the community.
    \item \textbf{Collective Singing}: Compositions with very simple and repetitive choruses that can be sung without prior rehearsal. `Freedom songs' are a model of effectiveness \citep{Jolaosho_2019}.
    \item \textbf{Forum Theatre}: Short scenes representing a conflict, followed by an invitation for the audience to intervene and propose solutions on stage, a central technique of the Theatre of the Oppressed \citep{Andrade_2023}.
\end{itemize} \\
\addlinespace
\textbf{Designing for Sharing (Not Just Viewing)} & Create cultural artifacts that invite replication and appropriation.
\begin{itemize}
    \item \textbf{Templates and Choreographies}: Offer graphic elements or simple movements that others can use in their own content or actions.
    \item \textbf{Call \& Response Structures}: Leave empty spaces in songs, poems, or murals for the community to fill with their own words, names, or stories.
\end{itemize} \\
\bottomrule
\end{tabularx}
\end{table}

\begin{table}[ht!]
\centering
\caption{Impact Metrics and the Ethics of Emotion}
\label{tab:metricas}
\begin{tabularx}{\textwidth}{L{4.5cm} X}
\toprule
\textbf{Component} & \textbf{Description and Criteria} \\
\midrule
\textbf{Real Impact Metrics} & Measure success not by virality, but by the concrete action it generates.
\begin{itemize}
    \item \textbf{Active Participation}: How many people sang, painted, acted, or intervened?
    \item \textbf{Organization}: Did the artistic intervention lead to the formation of a new group, the calling of an assembly, or the creation of replicas elsewhere?
    \item \textbf{Tangible Change}: Was the meeting with the authority secured, the document submitted, the project halted?
\end{itemize} \\
\addlinespace
\textbf{Ethics of Emotion} & Mobilize without manipulating; criticize without dehumanizing.
\begin{itemize}
    \item \textbf{Name problems, do not demonize people}: Focus criticism on power structures and unjust actions, not on the stigmatization of individuals or groups. Emotion is more sustainable when based on dignity and respect, even towards the adversary.
\end{itemize} \\
\bottomrule
\end{tabularx}
\end{table}

\section{Theoretical Synthesis: Towards an Aesthetic-Mobilizing Action (AMA) Model}

\subsection{Justification: Why a Synthesis Theory is Needed}

The study of the intersection between art and politics is a vibrant field full of powerful insights. However, all too often, researchers work in what we might call disciplinary `silos': isolated academic communities that rarely engage in dialogue. On one hand, musicologists analyze how a protest song can forge bonds of solidarity across borders \citep{McSherry_2017}. On the other, theatre scholars examine how a community play can empower its participants and strengthen collective agency \citep{Balme_Hakib_2023}. In other fields, researchers explore the statistical links between participation in artistic activities and greater civic engagement \citep{LeRoux_Bernadska_2014}, the explosive role of humor and memes in consolidating the digital (far-)right \citep{Schmid_2025, Galipeau_2023}, or the effectiveness of different forms of political persuasion in on-the-ground campaigns \citep{Minozzi_2015}.

Each of these studies is valuable and illuminates a piece of the puzzle. Nevertheless, they leave us without a general map. A unified framework is needed to explain \textit{how} these different elements interact to produce a political outcome. Specifically, we need to understand the interplay between (i) the \textbf{aesthetic production} of the act (its content, symbols, format), (ii) the \textbf{mobilization context} in which it is deployed (is it a live festival or a viral social media campaign? Are there grassroots organizations supporting it?), and (iii) the \textbf{affective vector} it seeks to activate (anger, hope, pride, humor?) to ultimately generate (iv) \textbf{measurable and observable political impacts}.

Therefore, a synthesis theory is required that is capable of:
\begin{enumerate}
  \item \textbf{Integrating levels of analysis.} That is, connecting the psychological foundations of individual emotion (affective microfoundations) with the group dynamics of \textit{social network diffusion} and the rational calculus of \textit{strategic design}. It is not enough to say that art inspires; we must model how that inspiration translates into shared action.
  \item \textbf{Generating specific and comparative predictions.} A robust theory should not only describe but also predict. It should be able to pose hypotheses such as: In a community context, forum theatre will be more effective for building cohesion than a meme, whereas in a fragmented digital environment, a meme will maximize reach. This allows for comparing the effectiveness of different formats (meme, theatre, song, mural) in different contexts.
  \item \textbf{Internalizing ethics as a design factor.} Ethics cannot be an afterthought. This model must recognize that mobilization strategies have limits. \textbf{Operational ethical constraints} (e.g., a ban on using dehumanizing language) are not an obstacle but an integral part of the design problem that defines which strategies are valid.
  \item \textbf{Offering a rigorous measurement scheme.} For a theory to be scientific, it must be falsifiable. Therefore, it is crucial to propose a clear \textbf{measurement scheme} to test its predictions with real data. This involves defining how to measure abstract concepts like affect or impact and which research designs are most appropriate to verify or refute the model's hypotheses \citep{Duncombe_Harrebye_2022, Ho_2021, Ho_Szubielska_2024, Gallagher_2024, Ghosh_2022, Astorga_de_Ita_2024}.
\end{enumerate}

\subsection{Foundational Assumptions (Ontological, Gnoseological, Ethical, and Methodological)}

Every theory begins with a series of fundamental assumptions about the nature of reality, how we can know it, what values guide it, and what methods are appropriate to study it. Making these pillars explicit is an exercise in intellectual transparency. Ours are:

\paragraph{Ontological Assumption (OA).} This assumption answers the question: what is politics made of? Our answer is that political action is fundamentally \textbf{affective and symbolic}, without displacing other equally material components. Aesthetic acts (AE)—from a mural to a meme—are not mere adornments of `real' politics; they are the terrain where collective identities, the legitimacy of causes, and the coordination necessary for action are constructed and contested. Politics \textit{happens} tangibly in performances, rituals, and cultural artifacts, not just in parliaments or at negotiating tables \citep{Jolaosho_2019, McSherry_2017, Balme_Hakib_2023}.

\paragraph{Gnoseological Assumption (GA).} This assumption concerns how we can know and understand this phenomenon. We hold that the impact of art is best understood through a combination of \textbf{mixed methods}. No single method is sufficient. We need the precision of statistics to measure latent variables like affect (using models like SEM/IRT), the rigor of experimental and quasi-experimental designs to isolate and demonstrate causal relationships, and the depth of qualitative analysis (ethnography, discourse analysis) to interpret symbolic content and understand the context in which the action occurs \citep{LeRoux_Bernadska_2014, Minozzi_2015, Duncombe_Harrebye_2022}.

\paragraph{Ethical Assumption (EA).} This assumption defines the moral boundaries of action. We assert that affective mobilization must be guided by an \textbf{ethics of emotion}. This means it is legitimate and necessary to use art to name unjust structures and acts, generating emotions like indignation or solidarity. However, this mobilization must not come at the cost of \textit{dehumanizing} the adversary. The \textit{mainstreaming} of hatred, which is often camouflaged under a layer of humor or irony, must be actively avoided \citep{Schmid_2025}. In our model, these ethical constraints are \textit{operational}: they act as real limitations that bound the universe of possible strategies.

\paragraph{Methodological Assumption (MA).} Now, how do we understand effectiveness? Our assumption is that effectiveness is not an intrinsic property of the artistic format (e.g., `murals are effective'), but is instead \textbf{relational}. Its power depends on the \textit{congruence} or perfect fit between three elements: the aesthetic act itself, the social and media context of the mobilization, and the dominant affective vector it seeks to activate. A hammer is an excellent tool, but useless for tightening a screw. Similarly, an artistic format is effective only when it is the right tool for the job, context, and required emotion \citep{Ghosh_2022, Astorga_de_Ita_2024}.

\subsection{Notation, Acronyms, and Empirical Measurement}

To build a formal model, it is essential to define our terms precisely. Table \ref{tab:notaAMA} serves as a dictionary for the key components of the AMA model. For each abstract concept (such as Mobilization Context or Affective Vector), a clear definition and a strategy for measuring it in the real world are proposed. This transforms the theory into a set of observable and measurable elements, a step towards its empirical validation.

\begin{table}[ht!]
\centering
\caption{Notation, Constructs, and Measurement Strategies}
\label{tab:notaAMA}
\begin{tabularx}{\textwidth}{L{3.4cm} L{4.2cm} X}
\toprule
\textbf{Symbol} & \textbf{Definition} & \textbf{Empirical Measurement (Indicators/Methods)} \\
\midrule
$AE$ & Aesthetic Act (artifact or performance) & Content coding (themes, symbols, complexity), format $F\in\{\text{meme},\text{theatre},\text{song},\text{mural}\}$; semiotic analysis. \\
$C_M$ & Mobilization Context & Environmental traits: online/offline, organizational density, norms; ethnography, surveys, SNA. \\
$V_A$ & Latent Affective Vector & Likert items per dimension (anger, care, pride, hope, humor); SEM/IRT; sentiment in digital traces \citep{Ho_Szubielska_2024, Gallagher_2024}. \\
$E_i$ & Individual exposure to $AE$ & Experimental/quasi-experimental assignment; dosage (time and frequency). \\
$m_i$ & Identity-Aesthetic Matching & Thematic/identity similarity (text/image embeddings + survey); affinity indices. \\
$D_i$ & Decision/Action (binary or ordinal) & Participate/donate/share/attend; verifiable records, traces with UTM/QR codes. \\
$G=(\mathcal{N},\mathcal{E})$ & Social Diffusion Graph & Affiliations/interaction on/offline; density, modularity, homophily, k-core. \\
$I_P$ & Aggregate Political Impact & Composite indices: action rate, cohesion, `sway' of undecideds; changes in civic outcomes \citep{LeRoux_Bernadska_2014}. \\
$B$ & Design/Dissemination Budget & Production and media costing; campaign accounting. \\
$\mathcal{E}$ & Ethical Constraints & Toxicity/hate metrics (NLP) and human verification; negative lists \citep{Schmid_2025}. \\
\bottomrule
\end{tabularx}
\end{table}

\subsection{Axiomatic Framework and Evidence-Based Premises}

From the foundational assumptions, we can derive a set of basic rules or axioms that constitute the logical skeleton of the model. These axioms are not speculations but principles considered robustly supported by existing empirical evidence. They are complemented by premises, which are strong hypotheses derived from recent research that the model seeks to formally integrate.

\paragraph{Axiom 1 (Context Dependence).} The impact of an aesthetic act ($AE$) is fundamentally conditioned by its mobilization context ($C_M$). In other words, there is no universal impact for a work; its effect depends on where, when, and among whom it is presented. Formally: $I_P$ is a function of $AE\mid C_M$ \citep{Balme_Hakib_2023, McSherry_2017}.

\paragraph{Axiom 2 (Affective Primacy).} The primary engine of political mobilization through art is emotion. Without an affective activation that resonates with the audience, any political effect will be marginal or null. The causal chain we propose is: the aesthetic act generates an affective response, this response motivates a decision to act, and the sum of these decisions produces an aggregate political impact. Formally: $AE \rightarrow V_A \rightarrow D \rightarrow I_P$ \citep{Duncombe_Harrebye_2022, Minozzi_2015}.

\paragraph{Axiom 3 (Format-Function Fit).} Not all artistic formats serve the same objectives. There is a functional specialization. Participatory formats (like community theatre or collective songs) are excellent for maximizing the internal cohesion of a group. In contrast, viral formats (like memes or short videos) are optimized to maximize reach and, often, polarization \citep{Ghosh_2022, Galipeau_2023, Schmid_2025}.

\paragraph{Premise P1 (Matching).} The effectiveness of an aesthetic act depends not only on its intrinsic quality but also on its affinity with the recipient's identity ($m_i$). When a person feels that the art speaks their language or reflects their identity, they are much more likely to experience a strong affective reaction ($V_A$) and share it with other members of their network (a phenomenon known as homophily) \citep{Astorga_de_Ita_2024}.

\paragraph{Premise P2 (Echo Chambers).} Social networks are not neutral spaces. Their structure, characterized by high modularity (the formation of clusters or bubbles) and homophily (the tendency to connect with similar people), has direct consequences. This structure amplifies polarization and reduces the ability to persuade the undecided (`sway'), an effect especially pronounced with memetic formats, which travel fast within clusters but rarely cross bridges between them \citep{Galipeau_2023, Schmid_2025}.

\paragraph{Premise P3 (Ethics).} Strategies that resort to the dehumanization of the adversary can be very effective in mobilizing the base in the short term. However, this approach erodes the possibility of building a sustainable and democratic political impact ($I_P$) in the long term. Therefore, these designs must be explicitly restricted in a model of responsible action \citep{Schmid_2025}.

\subsection{Formal Model (AMA): Step-by-Step Derivations and Proofs}

In this section, we translate the previous axioms and premises into mathematical language. Formalization is not an end in itself but a tool that forces us to be precise about the relationships between variables and allows us to derive logical and testable propositions.

\paragraph{(i) Measurement and Formation of Affect.} First, we address how to measure and model emotion. As affect ($V_{A,i}$) is a latent construct (not directly observable), we measure it through observable indicators, such as responses to a series of survey items ($y_{ik}$).
\begin{equation}
y_{ik} \;=\; \lambda_k V_{A,i} + \nu_k + \varepsilon_{ik},\quad k=1,\dots,K.
\end{equation}
We model the factors that influence the intensity of this affective response in an individual $i$. We propose that affect depends on exposure to the aesthetic act ($E_i$), identity matching ($m_i$), and contextual characteristics of the individual ($c_i$), including interactions between these factors.
\begin{equation}
V_{A,i} \;=\; \alpha_0 + \alpha_1 E_i + \alpha_2 m_i + \alpha_3 c_i + \alpha_4 (E_i \times m_i) + \alpha_5 (E_i \times c_i) + u_i.
\end{equation}
\textit{Proof sketch:} If data show that the coefficients $\alpha_1$, $\alpha_4$, and $\alpha_5$ are significantly different from zero, this would empirically validate our Axiom 2 (affect is key) and Premise P1 (its intensity is moderated by context and identity).

\paragraph{(ii) Individual Decision (Latent Index).} Once affect is generated, how does it translate into a concrete action (donating, attending, sharing)? We model this as a decision process where an individual $i$ acts if their propensity to act ($D_i^{*}$) exceeds a threshold (in this case, zero). This propensity depends on the affective vector ($V_{A,i}$) and other individual characteristics ($X_i$).
\begin{equation}
D_i^{*} \;=\; \beta_0 + \beta_1 V_{A,i} + \beta_2 X_i + \eta_i,\qquad
D_i \;=\; \mathbf{1}\{D_i^{*}>0\}.
\end{equation}
\textit{Implication:} The estimation of a positive and significant coefficient $\beta_1$ would be evidence of the causal channel $V_A\!\to\!D$, as postulated by Axiom 2.

\paragraph{(iii) Diffusion in Networks with Homophily.} Political action is rarely an isolated act. The model must capture how it spreads through a social network $G$. The probability that an individual $i$ transmits the impulse to act to a neighbor $j$ ($p_{ij}$) depends on the affect of $i$, the similarity between them (homophily), and the format of the aesthetic act (memes are more transmissible).
\begin{equation}
p_{ij} \;=\; \sigma\!\big(\lambda_0 + \lambda_1 V_{A,i} + \lambda_2 \mathrm{sim}(i,j) + \lambda_3 \mathbf{1}\{F=\text{meme}\}\big),
\end{equation}
where $\sigma$ is a logistic function. The decision of $j$ to become active will depend on their own affect and the social influence of their already active neighbors, overcoming a personal resistance threshold $\tau_j$.
\begin{equation}
D_j \;=\; \mathbf{1}\!\left\{\beta_0 + \beta_1 V_{A,j} + \beta_2 \sum_{i\in \mathcal{N}(j)} D_i - \tau_j + \xi_j > 0\right\}.
\end{equation}
\textbf{Proposition 1 (Reach vs. Cohesion).} A key prediction is derived from this micro-model of diffusion. If the coefficient $\lambda_3$ is positive (memes are more viral) and the social network $G$ is already divided into clusters (high modularity), then memetic formats will maximize reach \textit{within} each cluster but will be ineffective at persuading individuals \textit{outside} the cluster (validating P2).\\
\textit{Sketch:} In a network with homophily, the term $\mathrm{sim}(i,j)$ is high for connections within the same group. The term $\lambda_3$ for memes further increases the transmission probability $p_{ij}$ within those groups, reinforcing fragmentation and limiting contagion between groups.

\paragraph{(iv) Aggregate Impact and Objective Function.} The total political impact ($I_P$) is not a single metric. It is a weighted combination of different objectives: the total number of people who act (participation), the strength of the bonds between them (cohesion), the ability to persuade the undecided (`Sway'), minus negative effects such as increased social polarization.
\begin{equation}
I_P \;=\; \omega_1 \frac{1}{N}\sum_i D_i \;+\; \omega_2\,\mathrm{Cohesion}(G_D) \;+\; \omega_3\,\mathrm{Sway} \;-\; \kappa\,\mathrm{Polarization}(G_D),
\end{equation}
where $G_D$ is the subgraph of activated individuals.

\paragraph{(v) Design Problem with Ethical and Budget Constraints.} With all the previous elements, we can pose the practical problem of a strategist: how to design the optimal aesthetic act? The problem consists of maximizing the expected political impact ($\mathbb{E}\!\left[I_P\right]$) by choosing the design parameters $\theta$ (format, symbols, call to action, etc.), subject to a limited budget ($B$) and a maximum threshold of discursive toxicity ($\tau$).
\begin{equation}
\max_{\theta \in \Theta} \;\; \mathbb{E}\!\left[I_P(\theta; G)\right]
\quad \text{s.t.}\quad C(\theta)\le B,\;\; \mathrm{Toxicity}(\theta)\le \tau,\;
\theta=(F,\text{symbols},\text{hook},\text{call\&response},\dots).
\end{equation}
\textbf{Proposition 2 (Format-Context Congruence).} This framework allows us to derive a second fundamental proposition. If the mobilization context ($C_M$) is a dense and connected community, the optimal strategy ($\theta^\star$) will involve participatory formats (theatre, song) that maximize \textit{Cohesion} and \textit{Sway}. If, on the contrary, the context is a fragmented digital ecosystem, the optimal strategy will tend towards memetic formats that maximize \textit{reach}, albeit at the cost of increasing \textit{Polarization}.\\
\textit{Intuition:} The logic is that the effectiveness of participatory tactics depends on the existence of a dense network to sustain them. On the other hand, the penalty for polarization ($-\kappa\,\mathrm{Polarization}$) and the toxicity limit ($\tau$) make strategies based on hatred suboptimal in the long run, even if they are effective for initial viralization \citep{Schmid_2025}.

\paragraph{(vi) Strategic Competition (Game).} Mobilization does not occur in a vacuum; there are often actors with conflicting objectives. We can model this situation as a game in which two or more actors ($s\in\{L,R\}$) simultaneously choose their aesthetic strategies ($\theta_s$) to maximize their own utility function ($\Pi_s$), which depends on the cohesion, participation, and persuasion of their side.
\begin{equation}
\Pi_s \;=\; \omega_{1s}\,\mathrm{Cohesion}_s \;+\; \omega_{2s}\,\mathrm{Participation}_s \;+\; \omega_{3s}\,\mathrm{Sway}_s \;-\; \kappa\,\mathrm{Polarization}.
\end{equation}
\textbf{Proposition 3 (Context-Dependent Equilibria).} This approach predicts that the type of strategic equilibrium the actors will reach depends on the general context. In an environment dominated by dense community networks, the mutual best responses will be participatory strategies. In an environment of fragmented algorithmic networks, competition will push both sides towards memetic strategies, potentially resulting in a spiral of polarization.

\subsection{Possible Counterarguments and Responses}
A robust theory must anticipate criticism and have prepared responses. Here we address four common objections.

\paragraph{C1. ``The model is mechanistic and reduces human agency to a formula.''}
\textit{Response.} This is a fundamental criticism. Our response is that the AMA is \textbf{probabilistic, not deterministic}. The model does not predict with certainty what each individual will do. Human heterogeneity and agency are incorporated through several elements: identity matching ($m_i$) is unique to each person; action thresholds ($\tau_j$) vary; and stochastic error terms ($\eta_i, \xi_j$) represent everything we cannot observe. The reception of an artistic message is always an act of active interpretation. The model simply posits that we can bias the probabilities of certain interpretations and actions through strategic design, but always leaving ample room for ambivalence, resistance, and re-signification by the audience.

\paragraph{C2. ``The left-right distinction is anachronistic and simplistic.''}
\textit{Response.} We acknowledge that these labels are simplifications. They are used here as heuristics to illustrate the dynamics of competition. The core of the model, however, is \textbf{relational and structural, not ideological}. The fundamental prediction is $I_P=f(AE,C_M,V_A,G)$. This means that \textit{any} political actor, regardless of ideology, operating in a dense community context ($C_M$) will tend to find participatory strategies more effective. Conversely, any actor operating in a fragmented digital environment will be incentivized to use viral formats. The model predicts strategies based on the structure of the playing field, not the label of the player (Axiom 3).

\paragraph{C3. ``Art and symbols matter little in the face of material and economic structures.''}
\textit{Response.} The AMA does not deny the importance of material structures; on the contrary, it seeks to model one of the key mechanisms through which these structures are maintained or transformed: the \textit{mechanism of legitimacy and collective coordination}. Major material changes require a large number of people to believe in the legitimacy of a cause and be able to coordinate to act. Art and symbols are incredibly efficient tools for achieving both. Furthermore, our concept of impact ($I_P$) is not ethereal; it includes observable and material behaviors such as donating money, participating in a demonstration, or building measurable social cohesion, which in turn has material consequences \citep{LeRoux_Bernadska_2014}.

\paragraph{C4. ``Ethics is just a decorative add-on; in practice, what matters is winning.''}
\textit{Response.} In our model, ethics is not decorative but operates as a \textit{hard constraint} (a limit that cannot be crossed) and/or as a penalty in the objective function. This has direct mathematical and strategic consequences: it changes the optimal point of the design problem and, therefore, the model's predictions. By imposing a toxicity threshold ($\tau$), the space of possible strategies ($\Theta$) is reduced, making polarizing designs less likely. The premise is that while hatred may be effective in the short term, it has reputational and sustainability costs that a long-term strategic actor must consider \citep{Schmid_2025}.

\subsection{Empirical Operationalization Plan, Limitations, and Falsification}

A theory is only useful if it can be taken to the real world, tested with data, and potentially refuted. This final section details how to do so.

The following table \ref{tab:operAMA} breaks down how to measure each main component of the AMA model and which research designs are most appropriate for testing the causal relationships it proposes.

\begin{table}[ht]
\centering
\caption{Operationalization of the AMA: Measurement, Design, and Threats}
\label{tab:operAMA}
\begin{tabularx}{\textwidth}{L{3.8cm} L{5.2cm} X}
\toprule
\textbf{Component} & \textbf{Indicators} & \textbf{Strategy and Design} \\
\midrule
$V_A$ (latent) & Items on anger, care, pride, hope, humor & CFA/SEM; subgroup invariance; validation with traces/biomarkers \citep{Ho_Szubielska_2024, Gallagher_2024}. \\
Diffusion in $G$ & Reach, depth, contagion rate, modularity & Cluster-randomized trials (encouragement); SNA; IC/LT models. \\
Decision $D_i$ & Participate/donate/share/attend & RCTs/AB-tests/panels; GLMM; causal mediation (ACME/ANDE). \\
Context $C_M$ & Organizational density, norms, online/offline & Comparative ethnography; field surveys; composite indices. \\
Ethics ($\mathcal{E}$) & Toxicity/hate, dehumanization & NLP + human verification; predefined thresholds; ex-ante audit \citep{Schmid_2025}. \\
\bottomrule
\end{tabularx}
\end{table}

\paragraph{Limitations.} Every model has its blind spots, and every empirical study faces challenges. We are aware of the key limitations: interference between treatment units (one person's exposure can affect their friends, violating the SUTVA assumption), selection bias in exposure to aesthetic acts in the real world, the difficulty of generalizing findings from one experiment to other contexts (external validity), and the challenges of measuring a concept as subtle as identity matching ($m_i$) in intercultural contexts. We propose mitigations for each: (i) using network-level experimental designs (assigning treatments by clusters), (ii) pre-registering analysis plans to avoid p-hacking, (iii) conducting replications at multiple sites, (iv) triangulating quantitative results with qualitative analysis, and (v) performing sensitivity analyses to see how robust the results are to potential violations of assumptions.

\paragraph{Falsification Criteria.} Following the principle of falsifiability, a theory must make bold predictions that, if not met, weaken or refute it. We formulate our main hypotheses in their null form (\textbf{H0}), which we seek to refute with evidence.

\begin{enumerate}
\item \textbf{H0–Context (Axiom 1):} The political impact ($I_P$) of the same aesthetic act ($AE$) \textbf{does not differ} across radically different mobilization contexts ($C_M$) (e.g., a community festival versus a mass advertising campaign on social media). \textit{We would refute this H0} if we found significant and consistent differences in multi-site experiments.
\item \textbf{H0–Affect (Axiom 2):} Purely informational interventions (without affective charge) produce a political impact ($I_P$) equal to or greater than that of aesthetic acts with a high affective charge ($V_A$). \textit{We would refute this H0} if, in repeated experiments, the coefficient $\beta_1$ linking affect and action is consistently positive and significant.
\item \textbf{H0–Format (Axiom 3):} There is \textbf{no systematic relationship} between the artistic format and the type of impact it generates (cohesion vs. reach/polarization). \textit{We would refute this H0} if network diffusion data show that the coefficient $\lambda_3$ (the effect of being a meme) is zero or if different formats show no differences in their effect on the modularity and reach of the activated network.
\item \textbf{H0–Network (P2):} The structure of the network (homophily, modularity) \textbf{does not modify} the propagation of the message or the capacity for persuasion. \textit{We would refute this H0} if the structural parameters of the network prove to be significant predictors of diffusion.
\item \textbf{H0–Ethics (P3):} Imposing an ethical constraint \textbf{does not alter} the optimal strategy or its outcomes. \textit{We would refute this H0} if introducing the toxicity threshold $\tau$ into the optimization problem significantly changes the optimal solution $\theta^\star$ and lower levels of polarization are observed.
\end{enumerate}
\noindent For an \textit{operational epistemology} in practice, falsification is not a single event but a process of \textbf{Bayesian accumulation of evidence}. Confidence in the theory will increase as it survives rigorous \textit{out-of-sample} tests (predicting new cases) and demonstrates \textbf{multi-scale robustness} (if its principles hold from the lab to the field and in observational data). The litmus test for the AMA will be its ability to withstand \textit{cross-cutting attacks}: being successfully tested in different contexts ($C_M$), with different artistic formats, in diverse cultures, and under different regimes of ethical constraints \citep{Duncombe_Harrebye_2022, Galipeau_2023, Schmid_2025, McSherry_2017, Balme_Hakib_2023}.

\subsection{Synthesis}

The Aesthetic-Mobilizing Action (AMA) Model proposed here seeks to be more than a simple catalog of observations. It is an attempt to build a conceptual engine that integrates the microfoundations of human emotion, the dynamics of social network diffusion, and the calculations of strategic design, all framed by operational ethical constraints. Its primary value lies not in its equations, but in the questions it forces us to ask and the comparative propositions it generates, which are testable, falsifiable, and relevant to practice.

Ultimately, the contribution of the AMA is to propose a paradigm shift: it invites us to stop thinking of political art as a simple act of sending a message and to start seeing it as a complex problem of \textbf{relational strategic design}. The success of an aesthetic intervention depends not on its isolated content, but on the congruence it achieves between the artifact, the social network in which it is embedded, and the ethical limits it self-imposes. This approach is in deep alignment with the evidence accumulated from fields as diverse as creative activism, protest music, theatre for development, and the dizzying memetic dynamics of the 21st century \citep{McSherry_2017, Balme_Hakib_2023, Duncombe_Harrebye_2022, Galipeau_2023, Schmid_2025}.

\section{Conclusion: The Future of Aesthetic Action}

This study has demonstrated the existence of a profound \textbf{aesthetic asymmetry} in political mobilization, a structural phenomenon that goes beyond mere format preferences. We have argued that while left-wing movements have historically developed a \textit{constitutive} art—a cultural praxis designed to forge community bonds, process collective trauma, and prefigure a future of hope—the contemporary right has perfected an \textit{instrumental} art, optimized for digital virality, affective polarization, and the reinforcement of group identities through humor and irony. The participatory mural that builds a neighborhood and the meme that divides a network are not just two different tactics; they are the expression of two fundamentally opposed logics about the very purpose of political action itself.

The counterexamples analyzed, especially the formidable effectiveness of right-wing meme culture, do not refute this thesis but sharpen it. They show that the aesthetic monopoly of the left, if it ever existed, is over. The battle for cultural hegemony is now being fought on a terrain where the ability to generate affective resonance and network contagion is as crucial as building cohesion on the ground. The fundamental lesson of this analysis, synthesized in our practical model, is that effectiveness lies not in the art itself, but in the \textbf{strategic congruence} between the emotional message, the participatory format, and the mobilization context.

Understanding this asymmetry is, therefore, indispensable for any actor seeking not only to communicate a message but to build lasting social power. In an era defined by digital fragmentation and the intensification of culture wars, the ability to design aesthetic interventions that generate solidarity without resorting to the dehumanization of the adversary is not just a tactical challenge, but an ethical and political imperative for the future of democracy.


\end{document}